\documentclass[twocolumn,a4paper]{article}
\usepackage[spanish,es-nodecimaldot,es-tabla]{babel}
\usepackage[utf8]{inputenc}
\usepackage[T1]{fontenc}
\usepackage{apacite}
\usepackage{graphicx}
\usepackage{xcolor}
\usepackage{authblk}

\title{Simulación del flujo de viento sobre el territorio de Guatemala utilizando un modelo climático regional}

\author{Enrique Pazos\\epazos@ecfm.usac.edu.gt}
\affil{Instituto de Investigación, Escuela de Ciencias Físicas y Matemáticas,\\
Universidad de San Carlos de Guatemala}
\date{}

\begin{document}
\twocolumn[
\begin{@twocolumnfalse}
\maketitle
\begin{abstract}
Se presenta un análisis de la circulación del viento en la vecindad de la superficie
 terreste, para una región centrada alrededor de Guatemala. 
Para tal fin se utilizó el modelo climático regional (RegCM), con el cual se simuló la dinámica atmosférica sobre dicha región durante todo el 2016. 
El propósito del estudio es obtener la variación de mesoescala (decenas de kilómetros) del campo de velocidad del viento. 
Se puede observar que a medida que la resolución se incrementa se obtiene una representación más precisa de los detalles y características de la topografía del terreno, la cual influye en los patrones de circulación del viento a escalas de pocos kilómetros. 
Con una resolución fina de 2 km es posible observar zonas de flujo intenso de viento sobre la superficie; como en Palín, Escuintla. 
También se logra ver la presencia de patrones de circulación diurna que son producto del ciclo diario de calentamiento del terreno debido al sol y el consecuente enfriamiento durante la noche. Este es el primer reporte de una línea de estudio en donde se planea analizar las características climáticas propias de la región guatemalteca.\\
{\bf Palabras clave}: RegCM, resolución, simulación, energía eólica, topografía

-----

We present an analysis of the wind circulation in the vecinity of the ground surface, for a region centered around Guatemala. 
We used the regional climate model RegCM to simulate the atmospheric dynamics above that region during the full year 2016. 
The purpose of the study is to obtain the mesoscale variation (tens of kilometers) of the wind velocity field.
It can be seen that as resolution is increased, the details in topography are better represented, they in turn influence the wind circulation patterns on scales of a few kilometers. 
With a fine resolution of 2 km it is possible to confirm the existence of intense wind flux zones over the surface; such as Palín, Escuintla.
We are also able to observe diurnally varying circulations, which are the product of the daily cycle of terrain heating due to the sun and the subsecuent cooling during the night. This is the first report in a line of studies where we plan to analyze the climatic features of the Guatemalan region. 
\vspace{1cm}
\end{abstract}
\end{@twocolumnfalse}
]

\section{Introducción}
Una de las maneras de estudiar el clima es por medio de modelos numéricos computacionales. Aquellos modelos que toman como sistema de estudio toda la atmósfera del planeta se les llama Modelos de Circulación General (MCG). Éstos son útiles para analizar y prever el comportamiento y tendencia global de la atmósfera. Las resoluciones horizontales típicas utilizadas están alrededor de los 100 km~\cite{salathe2010regional,dosio2015dynamical}. Por otra parte, se encuentran los Modelos Climáticos Regionales (MCR) los cuales estudian la porción de la atmósfera que se encuentra sobre un dominio o región limitada de la superficie terrestre. Al ser un modelo de área limitada, un MCR posee fronteras laterales, en las cuales se utiliza la información generada por un MCG como condición de frontera. La ventaja de un MCR es que se pueden utilizar grillas de alta resolución horizontal para representar con fidelidad las características finas de la topografía del terreno. Un MCR con alta resolución es útil en estudios de adaptación e impacto ambiental~\cite{wang2004regional,lee2014potential}.

Cuando se utiliza un MCR se habla de su valor agregado. Esto se refiere a la ganancia en información acerca de las variables climáticas al utilizar una grilla horizontal con una resolución de decenas de kilómetros o menos. Dicha ganancia en información está representada por la variación espacial y/o temporal de los campos meteorológicos al incrementar la resolución por encima de la de los datos suministrados~\cite{lee2014potential}. En simulaciones donde el MCR incrementa la resolución de datos de reanálisis se ha reportado valor agregado en las variables de superficie, (temperatura y velocidad del viento a 10 m) en regiones caracterizadas por detalles orográficos en pequeña escala tales como regiones montañosas~\cite{di2012potential}. Al tener una mejor representación de las laderas de las montañas, el movimiento vertical de aire presenta un incremento adicional que no se observa en MCG. A mayor movimiento vertical, mayor es la precipitación dinámica, producida por el ascenso lento de masas de aire húmedo~\cite{wallace2006atmospheric}, que a su vez resulta en una atmósfera más seca y cálida~\cite{jones1995simulation,caldwell2010california,emanuel1999development}. 

Los MCR se han utilizado para estudiar la sensibilidad de los modelos a la resolución. Se ha encontrado que con una alta resolución en el MCR se obtiene menos precipitación ligera y más del tipo fuerte sobre regiones montañosas como los Alpes en Europa~\cite{giorgi1996investigation}, las cordilleras del estado de California~\cite{leung2003sensitivity} y las montañas del estado de Washington~\cite{salathe2010regional}. Otros estudios han tratado de encontrar mecanismos o explicaciones a fenómenos locales, tales como la canícula o veranillo en la región mesoamericana~\cite{small2007central,magana1999midsummer}. En la región de América del Sur se ha encontrado que la efectividad de un MCR depende de la parte de la Cordillera de los Andes que se estudie, siendo claves los análisis de sensibilidad para determinar las diferencias entre la información proporcionada por un MCR y un MCG~\cite{de2011assessing}. Los MCR también se han utilizado en estudios de evaluación de energía eólica y variabilidad del viento~\cite{garcia2013relationship}.

En el presente estudio se utiliza el MCR llamado RegCM~\cite{giorgi2012regcm4} para analizar la circulación horizontal del viento en la parte más baja de la capa de frontera planetaria, es decir, a pocos metros de altura sobre la superficie del terreno. El área de interés es el territorio guatemalteco, sobre el cual se utilizan dominios de extensión y resolución variable. El propósito de este trabajo es obtener el valor agregado de un MCR para los patrones de circulación del viento y el efecto que la topografía pueda tener sobre los mismos. En este caso, el valor agregado se encuentra en las variaciones del campo de velocidad a una escala menor que la de los datos iniciales. Dado que el valor de un campo en las celdas de la grilla representa el promedio de tal campo en el área que cubre, una mayor resolución proporciona valores promedio sobre áreas más pequeñas; obteniendo así información adicional sobre el patrón de flujo de viento.

\section{Materiales y métodos}
Los resultados que se presentan en este estudio fueron obtenidos con el MCR llamado RegCM versión 4.5, el cual ha sido desarrollado en el {\it Abdus Salam International Centre for Theoretical Physics} (ICTP). Se utiliza el núcleo no-hidrostático, el cual resuelve las ecuaciones de la dinámica de fluidos en las tres dimensiones espaciales. La topografía del terreno se incorpora en los cálculos mediante la coordenada vertical adimensional $\sigma$, la cual se define en términos de la presión atmosférica y la elevación del terreno~\cite{giorgi1996investigation}. La superficie de la tierra se representa con el valor constante $\sigma=1$, mientras que la frontera superior de la atmósfera adquiere el valor $\sigma=0$. RegCM utiliza el modelo de radiación {\it Community Climate Model} (CCM3) del {\it National Center for Atmospheric Research} (NCAR), el cual toma en cuenta el efecto de la presencia de gases como O$_3$, H$_2$O, CO$_2$ y O$_2$ en la atmósfera~\cite{kiehl1998national}. La interacción entre la parte inferior de la atmósfera, la vegetación y el contenido de humedad en el suelo se modela utilizando el {\it Biosphere-atmosphere transfer scheme} (BATS)~\cite{dickinson1993biosphere}. La precipitación convectiva sobre el suelo se calcula mediante el esquema de~\cite{grell1993prognostic} o el de Emanuel y \v{Z}ivkovi\'{c}-Rothman (1999).

Los resultados que se presentan se obtuvieron haciendo cinco corridas diferentes con RegCM. Para las primeras cuatro (V0-V3) el dominio en consideración es una región cuadrada, centrada en la república de Guatemala. Según se muestra en la Tabla~\ref{tab:runs}, éstas utilizan una resolución $\Delta s$ que empieza en 60 km y se refina hasta llegar a 2 km. El parámetro $\Delta s$ representa la separación entre los puntos de la grilla para ambas dimensiones horizontales. El dominio horizontal tiene una longitud de 1,800 km por lado para las corridas V0, V1 y V2. Para la corrida V3 el dominio tiene 960 km por lado. La reducción del tamaño del dominio obedece a la limitante impuesta por los recursos computacionales, lo cual se puede ver en el número de puntos que se utilizan para establecer la grilla horizontal. Utilizar más de $120\times120$ haría que el tiempo necesario para completar una simulación de un año tome más de dos semanas de ejecución ininterrumpida.  La Figura~\ref{fig:vmag} ilustra el área cubierta por el dominio de 960 km por lado. La corrida V4 utiliza una resolución $\Delta s=2$~km y cubre un área cuadrada de 120 km por lado en la región central de Guatemala, donde se ubican los volcanes de Acatenango, Fuego, Agua y Pacaya. La Figura~\ref{fig:vAcate} muestra el dominio utilizado en esta corrida.
\begin{table}
\begin{center}
\begin{tabular}{cccc}
\hline
corrida & no. de puntos & $\Delta s$ [km] & dominio [km] \\
\hline
V0	& 30 $\times$ 30   & 60 & 1,800 \\
V1	& 60 $\times$ 60   & 30 & 1,800 \\
V2	& 120 $\times$ 120 & 15 & 1,800 \\
V3	& 120 $\times$ 120 &  8 &  960 \\
V4      & 60 $\times$ 60   &  2 &  120 \\
\hline
\end{tabular}
\caption{Descripción del dominio computacional.}
\label{tab:runs}
\end{center}
\end{table}

Las corridas V0 a V3 abarcan un período de simulación del 1 de enero de 2016 hasta el 31 de diciembre del mismo año, mientras que en la corrida V4 la solución numérica se calcula solamente para los meses de noviembre y diciembre de 2016.

Los resultados de las simulaciones son almacenados en disco duro en el formato binario netCDF~\cite{rew1990netcdf}. Para las corridas que abarcan todo el 2016, la solución numérica se almacena cuando la simulación alcanza las 0, 6, 12 y 18 h, es decir que se tienen cuatro registros por día. Para la corrida V4, que comprende los últimos dos meses de 2016, el intervalo de salida de la simulación fue de 30 min, i.e. 48 registros por día. El objetivo de almacenar la solución con mayor frecuencia es poder tener una mejor representación en el tiempo de los procesos climáticos de período diurno.

Como en todo modelo climático regional, RegCM necesita condiciones iniciales y de frontera para poder resolver las ecuaciones diferenciales parciales de la dinámica de fluidos. En ambos casos se utilizaron los datos de reanálisis atmosférico global ERA-Interim, producidos por el {\it European Centre for Medium-Range Weather Forecasts} (ECMWF)~\cite{dee2011era}. El conjunto de datos ERA-Interim provee los valores de las variables dinámicas (temperatura, velocidad del viento, humedad, etc.) en función de las coordenadas longitud, latitud, altura y tiempo. La resolución en latitud y longitud es de 1.5$^\circ$, lo cual equivale a una separación en distancia de 167 km (en el ecuador). La resolución utilizada en todas las corridas es mayor a la de ERA-Interim, lo cual implica que los datos globales son interpolados en la grilla más fina que se utiliza en las simulaciones con RegCM.

Las condiciones de frontera lateral son impuestas a intervalos de seis horas. Esta es la información externa al dominio que dicta el comportamiento de la solución numérica sobre la región que se analiza. El resultado de la simulación es la combinación de la dinámica climática local (producto de la solución numérica de las ecuaciones), y de la influencia de los patrones climáticos globales que son alimentados al dominio a través de las fronteras laterales.

Las corridas V0, V1 y V2 utilizan los datos ERA-Interim para imponer las condiciones de frontera lateral. RegCM permite utilizar los datos de salida de una simulación como condición de frontera, siempre que el dominio esté enteramente contenido dentro del dominio de la solución que proveerá dicha condición de frontera. En este caso el dominio de V3 está contenido o anidado en el de V2, por lo que se utilizó la solución de la corrida V2 como datos de frontera para V3. La misma técnica fue empleada para darle condiciones de frontera a la corrida V4. La ventaja de usar una solución calculada previamente en lugar de los datos ERA-Interim como condición de frontera, es la incorporación de los patrones dinámicos locales que se desarrollan debido a una mejor representación tanto de la topografía del terreno como de la circulación del viento.

 En todas las corridas se utilizó 18 niveles verticales, con una presión atmosférica tope de 50 mb en la frontera superior. En este artículo se analizan únicamente las componentes horizontales de la velocidad del viento para el nivel vertical más cercano al suelo. Esto quiere decir que la altura para cual se muestran los resultados es de aproximadamente 20 m sobre el suelo.

El análisis de las variables climáticas requiere realizar el cálculo de múltiples operaciones y de promedios aritméticos, tanto en tiempo como en espacio. Para tal fin se utilizó el software llamado {\it netCDF Operator} (NCO)~\cite{zender2008analysis}, el cual consta de una serie de comandos de Linux para manipular archivos en formato binario netCDF.

\section{Resultados}
\subsection{Comparación con datos de estación}
\begin{figure}
\centering
\includegraphics[width=\columnwidth]{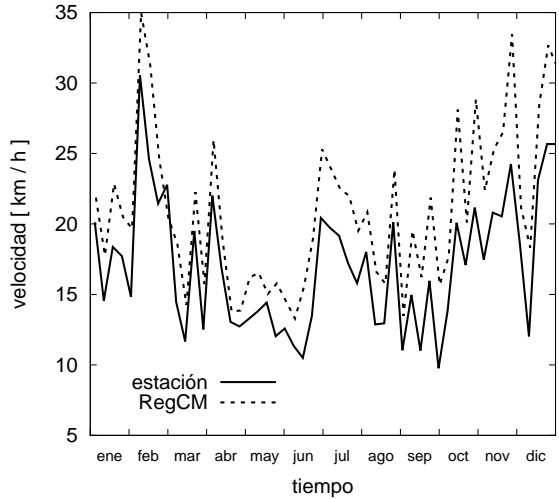}
\caption{Comparación de la magnitud de la velocidad del viento para diferentes resoluciones con datos de estación meteorológica. Los datos fueron obtenidos del sitio web del Insivumeh  para la estación La Aurora. Cada punto representa promedio semanal.}
\label{fig:LaAurora}  
\end{figure}

La Figura~\ref{fig:LaAurora} es una comparación de los resultados obtenidos de las corridas V0 a V3 y los datos recolectados por una estación meteorológica. 
La variable que se muestra es la magnitud de la velocidad horizontal del viento a lo largo del 2016 para la ubicación de la estación. Los datos que se utilizaron fueron recolectados por la estación La Aurora, perteneciente a la red de estaciones meteorológicas del Instituto Nacional de Sismología Vulcanología Meteorología e Hidrología (Insivumeh). La gráfica muestra los promedios semanales a lo largo de todo el año. Para cuantificar la cercanía de cada corrida a los datos recolectados se calculó la norma L1 y L2 de las diferencias entre los datos de estación y las corridas. La norma L1 da como resultado: 387, 285, 213 y 211; para las diferencias con las corridas V0, V1, V2 y V3; respectivamente. De forma similar, los resultados para la norma L2 son: 56.9, 41.9, 31.0 y 32.0 (ambas normas fueron truncadas a tres cifras significativas).

\subsection{Velocidad promedio}
\begin{figure}
\includegraphics[width=\columnwidth]{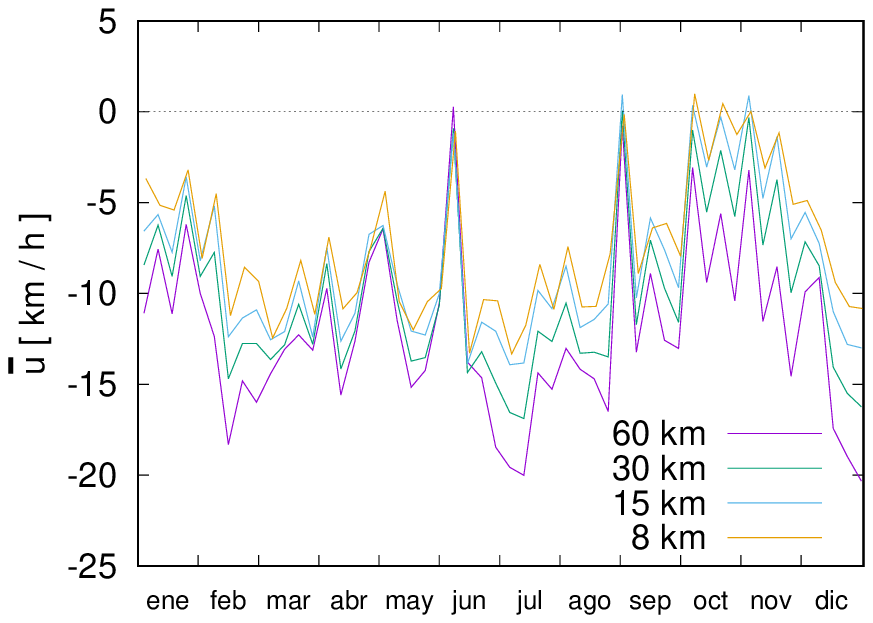}
\includegraphics[width=\columnwidth]{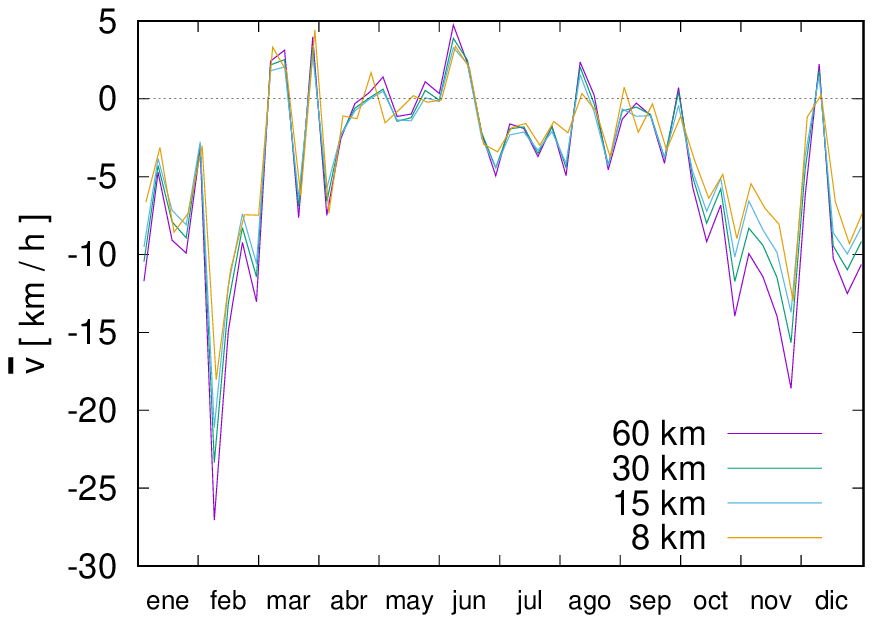}
\caption{Componente zonal $u$ y meridional $v$ del viento promediadas sobre todo el dominio para diferentes resoluciones $\Delta s$.}
\label{fig:uvAvg}
\end{figure}

\begin{figure*}
\centering
\includegraphics[width=\textwidth]{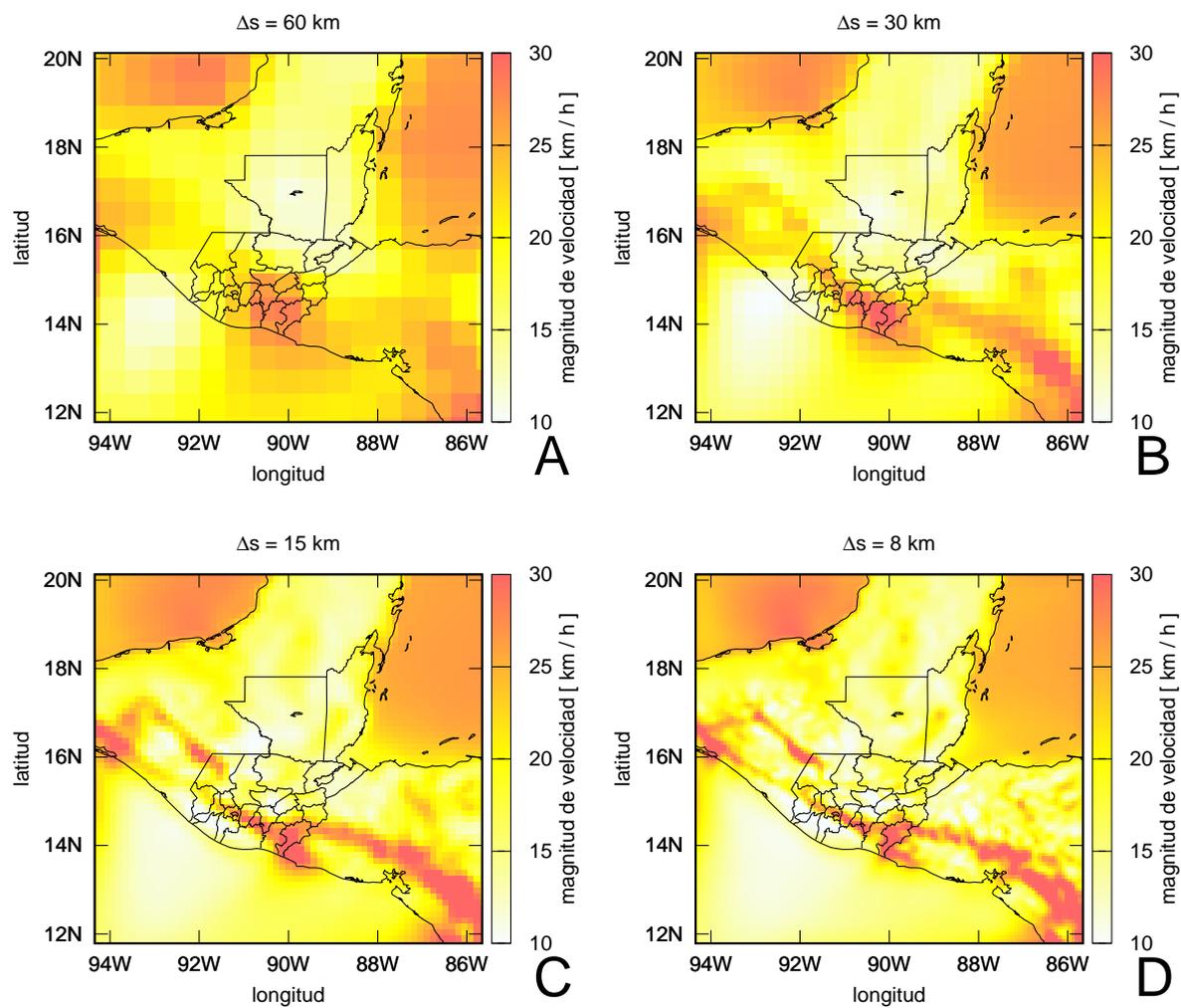}
\caption{Magnitud de la velocidad del viento promediada a lo largo de todo el 2016. Las resoluciones utilizadas son 60, 30, 15 y 8 km.}
\label{fig:vmag}
\end{figure*}

En la Figura~\ref{fig:uvAvg} se observan los promedios espaciales $\bar{u}$ y $\bar{v}$ de las componentes zonal $u$ y meridional $v$ del viento como función del tiempo. Los valores zonales positivos denotan dirección hacia el este y los negativos hacia el oeste, mientras que los valores meridionales positivos denotan dirección hacia el norte y los negativos hacia el sur.  El promedio se calcula sobre el dominio de la corrida V3 (mismo que se muestra en la Figura~\ref{fig:vmag}). Cada variable depende de cuatro coordenadas: longitud, latitud, altura y tiempo. El promedio de una variable sobre una región espacial se calcula como
\begin{equation}
\bar{u}(\sigma_k, t_n) = \frac{1}{N_xN_y} \sum_{i=1}^{N_x} \sum_{j=1}^{N_y} u(x_i,y_j,\sigma_k,t_n),
\end{equation}
donde $x_i$ e $y_j$ son las coordenadas longitud y latitud para la $ij$-ésima celda de la grilla, $\sigma_k$ es la altura del $k$-ésimo nivel vertical, $t_n$ es el tiempo a lo largo del año y $N_x$, $N_y$ es el número de puntos a lo largo de las coordenadas longitud, latitud. Las curvas de la gráfica corresponden a los diferentes valores de resolución $\Delta s$.

En la Figura~\ref{fig:vmag} se gráfica el valor promedio a lo largo de todo el 2016 de la magnitud de la velocidad horizontal. Los diferentes paneles muestran la misma variable para diferentes valores de la resolución espacial $\Delta s$. La magnitud del vector velocidad horizontal $V$ en términos de $u$ y $v$ se calcula como $V=\sqrt{u^2+v^2}$.  El promedio en el tiempo se calcula con el valor de $V$ en cada punto del dominio de acuerdo a
\begin{equation}
\langle V(x_i,y_j,\sigma_k) \rangle=\frac{1}{N_t} \sum_{n=1}^{N_t} V(x_i, y_j, \sigma_k, t_n),
\label{eq:promedio}
\end{equation}
donde $N_t$ es el número total de instantes en la dimensión temporal. Todos los casos de interés mostrados en este artículo se refieren al flujo del viento en las inmediaciones de la superficie terrestre, por lo cual $\sigma_k$ corresponde al nivel vertical de menor altura.

\begin{figure}
\centering
\includegraphics[width=\columnwidth]{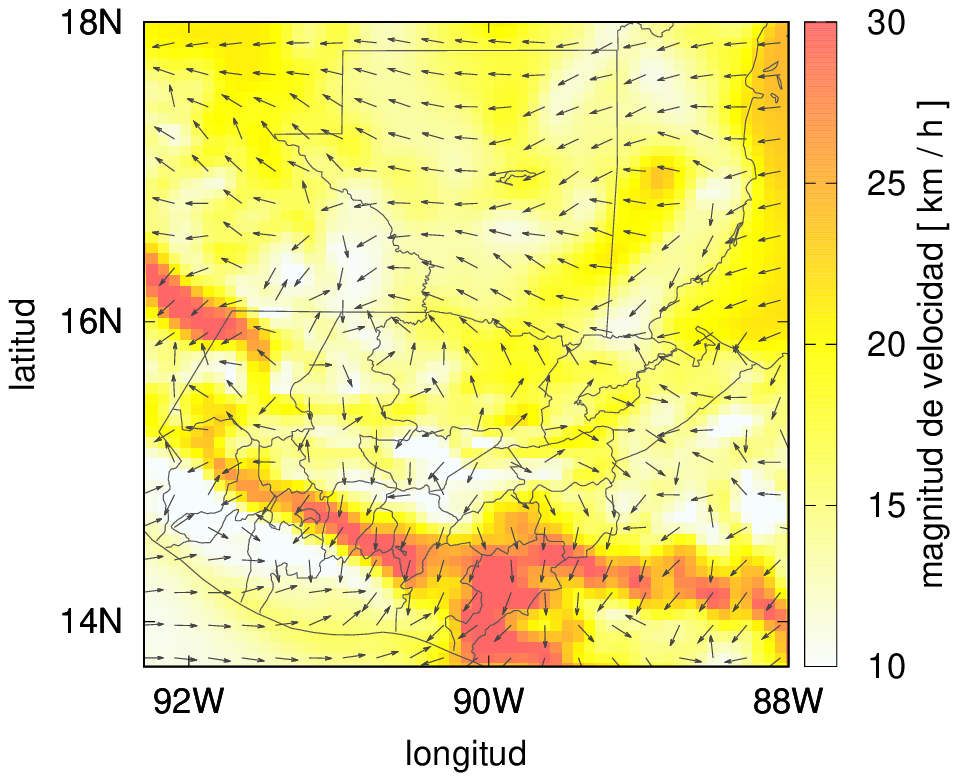}
\caption{Magnitud y dirección del viento promediada a lo largo de todo el año 2016. La resolución utilizada es de 8 km.}
\label{fig:vmagang3}
\end{figure}

En la Figura~\ref{fig:vmagang3} se muestra de nuevo la magnitud de la velocidad horizontal del viento promediada durante el 2016 y se agrega a la gráfica la dirección de la velocidad promediada a lo largo del mismo intervalo de tiempo. La dirección promedio del vector velocidad ${\bf V}$ se calcula como $\langle \theta \rangle = \tan^{-1}(\langle v \rangle/\langle u \rangle)$, para cada celda de la grilla. El valor promedio en el tiempo para $u$ y $v$ se obtiene con una fórmula similar a la Ec.~(\ref{eq:promedio}). En esta gráfica se ha hecho un acercamiento para mostrar más detalle sobre el territorio de Guatemala. Los resultados corresponden a la resolución $\Delta s=8$ km.

\begin{figure}
\centering
\includegraphics[width=\columnwidth]{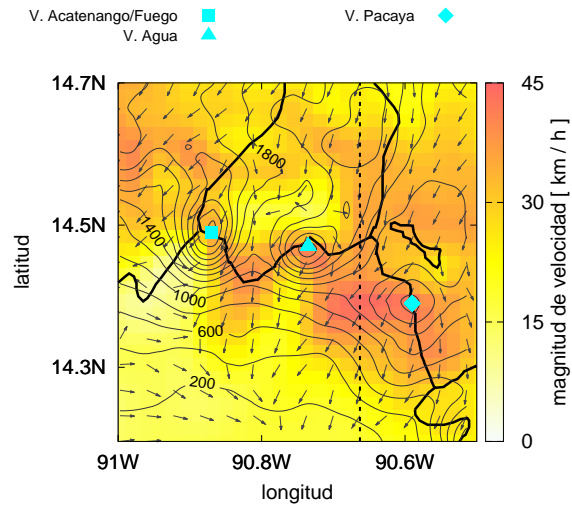}
\caption{Magnitud y dirección del viento promediada durante los meses de noviembre y diciembre de 2016. La resolución espacial es de 2 km. La elevación del terreno se muestra por medio de curvas de nivel etiquetadas por su altura en metros sobre el nivel del mar.}
\label{fig:vAcate}
\end{figure}

En la Figura~\ref{fig:vAcate} se muestra la magnitud y dirección del viento promediada únicamente en los meses de noviembre y diciembre de 2016. En esta corrida se utiliza la resolución más fina, con $\Delta s=$ 2 km. Esto representa un factor de 30 en el aumento de la resolución, comparada con el valor de la corrida V0, donde $\Delta s=60$~km. Con una resolución de 2 km es posible lograr una buena representación de la orografía en la región central de Guatemala. En la gráfica se identifica la ubicación de los volcanes de la región central: Acatenango, Fuego, Agua y Pacaya. Esta gráfica también muestra algunos contornos de nivel con la altura marcada en metros sobre el nivel del mar. Si bien con una grilla de 2 km de resolución es posible representar detalles salientes de la orografía, todavía no es suficiente para ver detalles finos, tal como la separación entre las cumbres de los volcanes de Fuego y Acatenango, que es de 3 km. Ambas formaciones se ven como un mismo volcán. Para esta corrida, las variables climáticas se almacenaron a intervalos de tiempo de 30 minutos. Esto hace que hayan 48 registros por día, lo cual da una mayor resolución temporal para el estudio de fenómenos climáticos con un período de duración de un día, tal como se presenta a continuación.

\subsection{Drenado de viento}
\begin{figure}
\centering
\includegraphics[width=\columnwidth]{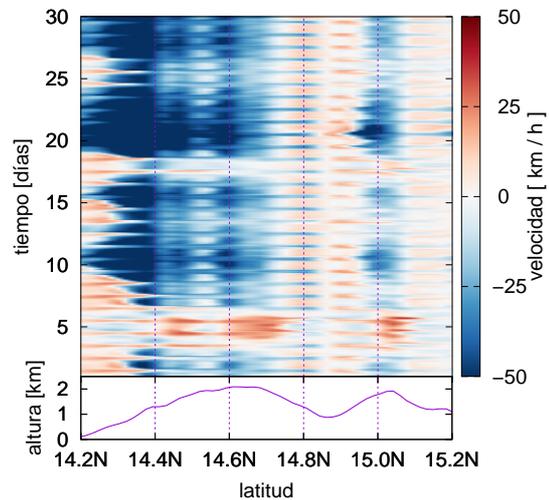}
\caption{Panel superior: componente meridional $v$ de la velocidad del viento para una sección norte-sur a una latitud fija de $90.6628^\circ$O. Esta sección corresponde a la línea vertical punteada que se muestra en la figura~\ref{fig:vAcate}. Panel inferior: perfil de elevación del suelo por donde pasa la línea punteada antes mencionada.}
\label{fig:vadrain}
\end{figure}

La Figura~\ref{fig:vadrain} consta de dos partes. El panel inferior muestra el perfil de elevación del suelo para un corte norte-sur a una longitud constante de $90.6628^\circ$O. Para colocar la escala de latitud en contexto geográfico, vale la pena mencionar algunos puntos conocidos con su respectiva latitud. Por ejemplo: la ciudad de Escuintla está a $14.3^\circ$N, la ciudad de Guatemala se encuentra a $14.6^\circ$N, el punto mínimo que aparece aproximadamente a $14.85^\circ$N es la cuenca del río Motagua y el límite territorial entre Baja Verapaz y Quiché se encuentra a una latitud alrededor de $15.2^\circ$N. El panel superior de la gráfica muestra la componente meridional $v$ del viento en función de la latitud y el tiempo. Viento hacia el norte es representado con grises claros y hacia el sur con grises oscuros. El tiempo está en unidades de días a partir de 1 de diciembre de 2016.


\section{Discusión}
A fin de comparar los resultados obtenidos de las simulaciones con mediciones reales, en la Figura~\ref{fig:LaAurora} se grafica el promedio semanal de la magnitud de la velocidad del viento obtenida con RegCM y medida por la estación meteorológica La Aurora del Insivumeh. Se puede ver que existe concordancia con la tendencia general, sin embargo el modelo sobreestima la velocidad del viento para todas las resoluciones. Esto podría ser un error sistemático tanto en la simulación como en el instrumento de medición. Sin tener información sobre la calibración del instrumento es difícil concluir acerca del origen de la diferencia. Si bien es cierto que RegCM sobreestima la velocidad del viento, también se aprecia que al aumentar la resolución las diferencias entre la simulación y los datos de estación es menor. Esto lo verifican las normas L1 y L2, cuyo valor decrece al disminuir $\Delta s$, excepto para L2 con la máxima resolución. Estos resultados indican que el cálculo de la velocidad del viento puede mejorar al aumentar la resolución, pero existe un límite más allá del cual el MCR ya no mejora la estimación.

Se puede observar en la Figura~\ref{fig:uvAvg} que la componente zonal $u$, es predominantemente negativa, lo que significa que el viento sopla hacia el oeste, el promedio de $u$ a lo largo del 2016 es de $-12.1$ km/h. 
De forma similar, el promedio anual de la componente meridional $v$ es de $-5.0$ km/h, lo cual indica que la dirección predominante es hacia el sur. 
En la componente meridional se observa que el viento sopla hace el sur durante los meses de enero, febrero y de octubre a diciembre. Durante el resto del año el viento sopla hacia el norte. 
Este patrón en la dirección del viento es lo que se observa empíricamente a lo largo del año.

La velocidad promedio mostrada en la Figura~\ref{fig:uvAvg} exhibe la tendencia de acercarse a cero a medida que se incrementa la resolución. 
Este efecto es más notorio para $\bar{u}$ que para $\bar{v}$, lo cual es entendible, pues se observa empíricamente que el viento tiende a soplar de norte a sur de manera constante durante la mayor parte del año. El efecto se acentúa para la componente zonal en los meses de junio, julio y agosto; donde $|\bar{u}|$ alcanza sus valores máximos en la dirección oeste, sobre todo para la resolución $\Delta s=60$~km. Este comportamiento se explica al tomar en cuenta que un $\Delta s$ menor (i.e. mayor resolución) implica una representación más fiel de la orografía (montañas) e irregularidades del terreno, cualidades que constituyen un obstáculo al flujo libre del viento. La diferencia es mínima en los picos de velocidad máxima, los cuales son influenciados en menor medida por el terreno y más por patrones de circulación sinóptica como ciclones y/o sistemas de alta y baja presión, originados más allá del dominio en consideración. Como consecuencia, el viento tiende a ser frenado, con la obvia implicación que $|\bar{u}|$ y $|\bar{v}|$ tienden a ser menores. La influencia de la elevación del terreno sobre el flujo del viento se verifica también al analizar las capas superiores de la atmósfera, en donde la velocidad permanece prácticamente invariante al cambiar la resolución de la grilla.

En la Figura~\ref{fig:vmag} se puede observar el aumento de detalle al aumentar la resolución. El panel A utiliza $\Delta s=60$ km. Es evidente que a este nivel de detalle no se tiene suficiente información sobre el territorio guatemalteco. Una sola celda de la grilla es suficiente para cubrir departamentos como Sacatepequez, Sololá o Totonicapán. Todo el detalle de la orografía y su influencia sobre la circulación local del viento está ausente. En los paneles B y C se utilizan resoluciones de 30 y 15 km, respectivamente. Se empieza a ver una distinción con más detalle de regiones de alta y baja velocidad. La característica saliente es que existe una franja de alta velocidad que coincide geográficamente con la Sierra Madre que atraviesa el territorio de Guatemala de oeste a este, siguiendo el perfil de la costa sur. En el panel D la resolución utilizada es de 8 km. Se puede observar que las regiones con velocidades más altas corresponden a las laderas montañosas de la Sierra Madre y la parte sur del departamento de Jutiapa. Los resultados mostrados en la Figura~\ref{fig:vmag} ilustran la ventaja de utilizar un modelo climático regional, que es la capacidad de observar la influencia de los accidentes geográficos de mesoescala en el comportamiento promedio del flujo de viento. La información acerca del flujo del viento se completa en la Figura~\ref{fig:vmagang3}, en la cual se ha agregado también la dirección promedio del viento. La franja de alta velocidad que coincide con la Sierra Madre tiene una dirección predominante hacia el sur.

Es posible alcanzar todavía mayor detalle en la simulación de la dinámica atmosférica siempre que hayan características del terreno que aparezcan al aumentar la resolución. La Figura~\ref{fig:vAcate} muestra tal situación. Aquí se ha utilizado un $\Delta s=2$ km, con lo cual se hacen evidentes las formaciones volcánicas más prominentes de la región central de Guatemala. Se puede apreciar que la dirección promedio en noviembre y diciembre es hacia el sur. El valor promedio máximo de la velocidad se alcanza en la región que se ubica entre los volcanes de Agua y Pacaya, que corresponde al municipio de Palín en el departamento de Escuintla. En la gráfica se puede apreciar que en esta región la dirección promedio es prácticamente perpendicular a las curvas de nivel del terreno. Esta peculiaridad se explica como un drenado de viento proveniente del área de la ciudad de Guatemala, la cual se encuentra a mayor altura. El viento que desciende desde el valle central tiende a acelerarse al bajar por el gradiente de altura que pasa por Palín. Un efecto similar, pero de menor proporción se observa entre los volcanes de Acatenango y Agua, donde la dirección promedio del viento también es perpendicular a las curvas de nivel del terreno. Si bien dos meses de simulación no son tiempo suficiente para considerar estos resultados como una característica climatológica estable, es un indicativo del potencial de estas regiones para la extracción de energía del viento. No es coincidencia que ya existe un parque eólico en las inmediaciones de Palín.

Utilizando una resolución de 2 km y registros de la velocidad del viento a cada media hora, es posible observar fenómenos de circulación térmica diurna, tal efecto se muestra en la Figura~\ref{fig:vadrain}.  Durante un día de buen tiempo, el sol calienta la pendiente de las montañas elevando su temperatura. El aire tiende a ascender a la largo de la pendiente, creando viento anabático. Durante la noche el terreno se enfría y la circulación se invierte, es decir, el aire baja por la pendiente estableciendo un viento katabático~\cite{wallace2006atmospheric}. Las latitudes 14.8$^\circ$N y 14.9$^\circ$N son un ejemplo perfecto de la circulación anabática y katabática. 
Estas latitudes corresponden a puntos que están en los lados sur y norte (a la misma longitud) de la cuenca del río Motagua, respectivamente. 
Se puede apreciar que al mismo tiempo que el viento sopla hacia el sur en el lado sur de la cuenca, éste sopla hacia el norte en el lado norte de la cuenca. 
En otras palabras, el viento asciende en ambas laderas. Durante la noche la dirección de la circulación se invierte. 
Es notorio que la franja de tiempo de circulación anabática es más corta que la katabática. 
Un patrón similar puede observase en la latitud 14.2$^\circ$N, donde los días de buen tiempo llegan hasta el 9 de diciembre y luego se observa fuerte viento proveniente del norte. 
En este caso, en los primeros días del mes no se alcanza una velocidad hacia el sur, el patrón de circulación es más bien un aumento y disminución de la velocidad hacia el norte, el cual tiene también un patrón diurno. En los días alrededor del 20 de diciembre el patrón se rompe debido a la entrada de un frente frío que hace que el viento sople hacia el sur.


En conclusión, se ha utilizado un MCR para realizar corridas de alta resolución sobre el territorio guatemalteco. El valor agregado que se puede extraer es la variación del campo de velocidad a una escala menor que la de los datos que proveen las condiciones iniciales y de frontera. En este caso el estudio ha sido enfocado a los patrones de circulación del viento en las inmediaciones de la superficie terrestre. Al contrario de lo que sucede a alturas superiores de la atmósfera donde el viento es geostrófico, la circulación del viento en la capa de frontera atmosférica está influenciada por la orografía, la textura del suelo y la presencia de ciudades. Se ha observado que la representación detallada de la topografía por medio de grillas de alta resolución influye marcadamente en la circulación del viento a mesoescala. Aunque el incremento en resolución ha permitido mejorar la exactitud de la velocidad del viento, la norma de las diferencias entre los datos de estación (para un punto) y las simulaciones indican que hay una resolución más allá de la cual los cálculos ya no mejoran.
Uno de los factores importantes del estudio y análisis de los patrones de circulación del viento es la creciente demanda de producción de energía limpia. Poder identificar zonas con un flujo fuerte y constante de viento es clave para el aprovechamiento de la energía eólica.

En trabajos futuros se analizarán las variables de precipitación y temperatura, las cuales tienen un efecto visible e inmediato sobre las diferentes actividades humanas.

\section{Agradecimientos}
A Vittorio M. Canuto del NASA Goddard Institute for Space Studies por haber presentado la ciencia del clima como un campo emocionante para investigar.

\bibliographystyle{apacite}
\bibliography{references-clima}

\begin{thebibliography}{}

\bibitem [\protect \citeauthoryear {%
Caldwell%
}{%
Caldwell%
}{%
{\protect \APACyear {2010}}%
}]{%
caldwell2010california}
\APACinsertmetastar {%
caldwell2010california}%
\begin{APACrefauthors}%
Caldwell, P.%
\end{APACrefauthors}%
\unskip\
\newblock
\APACrefYearMonthDay{2010}{}{}.
\newblock
{\BBOQ}\APACrefatitle {California wintertime precipitation bias in regional and
  global climate models} {California wintertime precipitation bias in regional
  and global climate models}.{\BBCQ}
\newblock
\APACjournalVolNumPages{Journal of Applied Meteorology and
  Climatology}{49}{10}{2147--2158}.
\newblock
\begin{APACrefDOI} \doi{10.1175/2010JAMC2388.1} \end{APACrefDOI}
\PrintBackRefs{\CurrentBib}

\bibitem [\protect \citeauthoryear {%
Dee%
\ \protect \BOthers {.}}{%
Dee%
\ \protect \BOthers {.}}{%
{\protect \APACyear {2011}}%
}]{%
dee2011era}
\APACinsertmetastar {%
dee2011era}%
\begin{APACrefauthors}%
Dee, D\BPBI P.%
, Uppala, S\BPBI M.%
, Simmons, A\BPBI J.%
, Berrisford, P.%
, Poli, P.%
, Kobayashi, S.%
\BDBL {}Vitart, F.%
\end{APACrefauthors}%
\unskip\
\newblock
\APACrefYearMonthDay{2011}{}{}.
\newblock
{\BBOQ}\APACrefatitle {The ERA-Interim reanalysis: Configuration and
  performance of the data assimilation system} {The era-interim reanalysis:
  Configuration and performance of the data assimilation system}.{\BBCQ}
\newblock
\APACjournalVolNumPages{Quarterly Journal of the Royal Meteorological
  Society}{137}{656}{553--597}.
\newblock
\begin{APACrefDOI} \doi{10.1002/qj.828} \end{APACrefDOI}
\PrintBackRefs{\CurrentBib}

\bibitem [\protect \citeauthoryear {%
De~Sales%
\ \BBA {} Xue%
}{%
De~Sales%
\ \BBA {} Xue%
}{%
{\protect \APACyear {2011}}%
}]{%
de2011assessing}
\APACinsertmetastar {%
de2011assessing}%
\begin{APACrefauthors}%
De~Sales, F.%
\BCBT {}\ \BBA {} Xue, Y.%
\end{APACrefauthors}%
\unskip\
\newblock
\APACrefYearMonthDay{2011}{}{}.
\newblock
{\BBOQ}\APACrefatitle {Assessing the dynamic-downscaling ability over South
  America using the intensity-scale verification technique} {Assessing the
  dynamic-downscaling ability over south america using the intensity-scale
  verification technique}.{\BBCQ}
\newblock
\APACjournalVolNumPages{International Journal of
  Climatology}{31}{8}{1205--1221}.
\newblock
\begin{APACrefDOI} \doi{10.1002/joc.2139} \end{APACrefDOI}
\PrintBackRefs{\CurrentBib}

\bibitem [\protect \citeauthoryear {%
Dickinson%
, Kennedy%
\BCBL {}\ \BBA {} Henderson-Sellers%
}{%
Dickinson%
\ \protect \BOthers {.}}{%
{\protect \APACyear {1993}}%
}]{%
dickinson1993biosphere}
\APACinsertmetastar {%
dickinson1993biosphere}%
\begin{APACrefauthors}%
Dickinson, R\BPBI E.%
, Kennedy, P.%
\BCBL {}\ \BBA {} Henderson-Sellers, A.%
\end{APACrefauthors}%
\unskip\
\newblock
\APACrefYear{1993}.
\newblock
\APACrefbtitle {Biosphere-atmosphere transfer scheme (BATS) version 1e as
  coupled to the NCAR community climate model} {Biosphere-atmosphere transfer
  scheme (bats) version 1e as coupled to the ncar community climate model}.
\newblock
\APACaddressPublisher{Boulder, Colorado}{National Center for Atmospheric
  Research, Climate {and} Global Dynamics Division}.
\newblock
\begin{APACrefDOI} \doi{10.5065/D67W6959} \end{APACrefDOI}
\PrintBackRefs{\CurrentBib}

\bibitem [\protect \citeauthoryear {%
Di~Luca%
, de Elía%
\BCBL {}\ \BBA {} Laprise%
}{%
Di~Luca%
\ \protect \BOthers {.}}{%
{\protect \APACyear {2012}}%
}]{%
di2012potential}
\APACinsertmetastar {%
di2012potential}%
\begin{APACrefauthors}%
Di~Luca, A.%
, de Elía, R.%
\BCBL {}\ \BBA {} Laprise, R.%
\end{APACrefauthors}%
\unskip\
\newblock
\APACrefYearMonthDay{2012}{}{}.
\newblock
{\BBOQ}\APACrefatitle {Potential for added value in precipitation simulated by
  high-resolution nested regional climate models and observations} {Potential
  for added value in precipitation simulated by high-resolution nested regional
  climate models and observations}.{\BBCQ}
\newblock
\APACjournalVolNumPages{Climate Dynamics}{38}{5-6}{1229--1247}.
\newblock
\begin{APACrefDOI} \doi{10.1007/s00382-011-1068-3} \end{APACrefDOI}
\PrintBackRefs{\CurrentBib}

\bibitem [\protect \citeauthoryear {%
Dosio%
, Panitz%
, Schubert-Frisius%
\BCBL {}\ \BBA {} L{\"u}thi%
}{%
Dosio%
\ \protect \BOthers {.}}{%
{\protect \APACyear {2015}}%
}]{%
dosio2015dynamical}
\APACinsertmetastar {%
dosio2015dynamical}%
\begin{APACrefauthors}%
Dosio, A.%
, Panitz, H\BHBI J.%
, Schubert-Frisius, M.%
\BCBL {}\ \BBA {} L{\"u}thi, D.%
\end{APACrefauthors}%
\unskip\
\newblock
\APACrefYearMonthDay{2015}{}{}.
\newblock
{\BBOQ}\APACrefatitle {Dynamical downscaling of CMIP5 global circulation models
  over CORDEX-Africa with COSMO-CLM: evaluation over the present climate and
  analysis of the added value} {Dynamical downscaling of cmip5 global
  circulation models over cordex-africa with cosmo-clm: evaluation over the
  present climate and analysis of the added value}.{\BBCQ}
\newblock
\APACjournalVolNumPages{Climate Dynamics}{44}{9-10}{2637--2661}.
\newblock
\begin{APACrefDOI} \doi{10.1007/s00382-014-2262-x} \end{APACrefDOI}
\PrintBackRefs{\CurrentBib}

\bibitem [\protect \citeauthoryear {%
Emanuel%
\ \BBA {} {\v{Z}}ivkovi{\'c}-Rothman%
}{%
Emanuel%
\ \BBA {} {\v{Z}}ivkovi{\'c}-Rothman%
}{%
{\protect \APACyear {1999}}%
}]{%
emanuel1999development}
\APACinsertmetastar {%
emanuel1999development}%
\begin{APACrefauthors}%
Emanuel, K\BPBI A.%
\BCBT {}\ \BBA {} {\v{Z}}ivkovi{\'c}-Rothman, M.%
\end{APACrefauthors}%
\unskip\
\newblock
\APACrefYearMonthDay{1999}{}{}.
\newblock
{\BBOQ}\APACrefatitle {Development and evaluation of a convection scheme for
  use in climate models} {Development and evaluation of a convection scheme for
  use in climate models}.{\BBCQ}
\newblock
\APACjournalVolNumPages{Journal of the Atmospheric
  Sciences}{56}{11}{1766--1782}.
\newblock
\begin{APACrefDOI} \doi{10.1175/1520-0469(1999)056<1766:DAEOAC>2.0.CO;2}
  \end{APACrefDOI}
\PrintBackRefs{\CurrentBib}

\bibitem [\protect \citeauthoryear {%
García-Bustamante%
\ \protect \BOthers {.}}{%
García-Bustamante%
\ \protect \BOthers {.}}{%
{\protect \APACyear {2013}}%
}]{%
garcia2013relationship}
\APACinsertmetastar {%
garcia2013relationship}%
\begin{APACrefauthors}%
García-Bustamante, E.%
, González-Rouco, J\BPBI F.%
, Navarro, J.%
, Xoplaki, E.%
, Luterbacher, J.%
, Jiménez, P\BPBI A.%
\BDBL {}Lucio-Eceiza, E\BPBI E.%
\end{APACrefauthors}%
\unskip\
\newblock
\APACrefYearMonthDay{2013}{}{}.
\newblock
{\BBOQ}\APACrefatitle {Relationship between wind power production and {N}orth
  {A}tlantic atmospheric circulation over the northeastern {I}berian
  {P}eninsula} {Relationship between wind power production and {N}orth
  {A}tlantic atmospheric circulation over the northeastern {I}berian
  {P}eninsula}.{\BBCQ}
\newblock
\APACjournalVolNumPages{{C}limate {D}ynamics}{40}{3-4}{935--949}.
\newblock
\begin{APACrefDOI} \doi{10.1007/s00382-012-1451-8} \end{APACrefDOI}
\PrintBackRefs{\CurrentBib}

\bibitem [\protect \citeauthoryear {%
Giorgi%
\ \protect \BOthers {.}}{%
Giorgi%
\ \protect \BOthers {.}}{%
{\protect \APACyear {2012}}%
}]{%
giorgi2012regcm4}
\APACinsertmetastar {%
giorgi2012regcm4}%
\begin{APACrefauthors}%
Giorgi, F.%
, Coppola, E.%
, Solmon, F.%
, Mariotti, L.%
, Sylla, M\BPBI B.%
, Bi, X.%
\BDBL {}Brankovic, C.%
\end{APACrefauthors}%
\unskip\
\newblock
\APACrefYearMonthDay{2012}{}{}.
\newblock
{\BBOQ}\APACrefatitle {RegCM4: {M}odel description and preliminary tests over
  multiple {CORDEX} domains} {Regcm4: {M}odel description and preliminary tests
  over multiple {CORDEX} domains}.{\BBCQ}
\newblock
\APACjournalVolNumPages{Climate Research}{52}{}{7--29}.
\newblock
\begin{APACrefDOI} \doi{10.3354/cr01018} \end{APACrefDOI}
\PrintBackRefs{\CurrentBib}

\bibitem [\protect \citeauthoryear {%
Giorgi%
\ \BBA {} Marinucci%
}{%
Giorgi%
\ \BBA {} Marinucci%
}{%
{\protect \APACyear {1996}}%
}]{%
giorgi1996investigation}
\APACinsertmetastar {%
giorgi1996investigation}%
\begin{APACrefauthors}%
Giorgi, F.%
\BCBT {}\ \BBA {} Marinucci, M\BPBI R.%
\end{APACrefauthors}%
\unskip\
\newblock
\APACrefYearMonthDay{1996}{}{}.
\newblock
{\BBOQ}\APACrefatitle {A investigation of the sensitivity of simulated
  precipitation to model resolution and its implications for climate studies}
  {A investigation of the sensitivity of simulated precipitation to model
  resolution and its implications for climate studies}.{\BBCQ}
\newblock
\APACjournalVolNumPages{Monthly Weather Review}{124}{1}{148--166}.
\newblock
\begin{APACrefDOI} \doi{10.1175/1520-0493(1996)124<0148:AIOTSO>2.0.CO;2}
  \end{APACrefDOI}
\PrintBackRefs{\CurrentBib}

\bibitem [\protect \citeauthoryear {%
Grell%
}{%
Grell%
}{%
{\protect \APACyear {1993}}%
}]{%
grell1993prognostic}
\APACinsertmetastar {%
grell1993prognostic}%
\begin{APACrefauthors}%
Grell, G\BPBI A.%
\end{APACrefauthors}%
\unskip\
\newblock
\APACrefYearMonthDay{1993}{}{}.
\newblock
{\BBOQ}\APACrefatitle {Prognostic evaluation of assumptions used by cumulus
  parameterizations} {Prognostic evaluation of assumptions used by cumulus
  parameterizations}.{\BBCQ}
\newblock
\APACjournalVolNumPages{Monthly Weather Review}{121}{3}{764--787}.
\newblock
\begin{APACrefDOI} \doi{10.1175/1520-0493(1993)121<0764:PEOAUB>2.0.CO;2}
  \end{APACrefDOI}
\PrintBackRefs{\CurrentBib}

\bibitem [\protect \citeauthoryear {%
Jones%
, Murphy%
\BCBL {}\ \BBA {} Noguer%
}{%
Jones%
\ \protect \BOthers {.}}{%
{\protect \APACyear {1995}}%
}]{%
jones1995simulation}
\APACinsertmetastar {%
jones1995simulation}%
\begin{APACrefauthors}%
Jones, R\BPBI G.%
, Murphy, J\BPBI M.%
\BCBL {}\ \BBA {} Noguer, M.%
\end{APACrefauthors}%
\unskip\
\newblock
\APACrefYearMonthDay{1995}{}{}.
\newblock
{\BBOQ}\APACrefatitle {Simulation of climate change over europe using a nested
  regional-climate model. {I}: {A}ssessment of control climate, including
  sensitivity to location of lateral boundaries} {Simulation of climate change
  over europe using a nested regional-climate model. {I}: {A}ssessment of
  control climate, including sensitivity to location of lateral
  boundaries}.{\BBCQ}
\newblock
\APACjournalVolNumPages{Quarterly Journal of the Royal Meteorological
  Society}{121}{526}{1413--1449}.
\newblock
\begin{APACrefDOI} \doi{10.1002/qj.49712152610} \end{APACrefDOI}
\PrintBackRefs{\CurrentBib}

\bibitem [\protect \citeauthoryear {%
Kiehl%
\ \protect \BOthers {.}}{%
Kiehl%
\ \protect \BOthers {.}}{%
{\protect \APACyear {1998}}%
}]{%
kiehl1998national}
\APACinsertmetastar {%
kiehl1998national}%
\begin{APACrefauthors}%
Kiehl, J.%
, Hack, J.%
, Bonan, G.%
, Boville, B.%
, Williamson, D.%
\BCBL {}\ \BBA {} Rasch, P.%
\end{APACrefauthors}%
\unskip\
\newblock
\APACrefYearMonthDay{1998}{}{}.
\newblock
{\BBOQ}\APACrefatitle {The national center for atmospheric research community
  climate model: {CCM}3} {The national center for atmospheric research
  community climate model: {CCM}3}.{\BBCQ}
\newblock
\APACjournalVolNumPages{Journal of Climate}{11}{6}{1131--1149}.
\newblock
\begin{APACrefDOI} \doi{10.1175/1520-0442(1998)011<1131:TNCFAR>2.0.CO;2}
  \end{APACrefDOI}
\PrintBackRefs{\CurrentBib}

\bibitem [\protect \citeauthoryear {%
Lee%
\ \BBA {} Hong%
}{%
Lee%
\ \BBA {} Hong%
}{%
{\protect \APACyear {2014}}%
}]{%
lee2014potential}
\APACinsertmetastar {%
lee2014potential}%
\begin{APACrefauthors}%
Lee, J.%
\BCBT {}\ \BBA {} Hong, S.%
\end{APACrefauthors}%
\unskip\
\newblock
\APACrefYearMonthDay{2014}{}{}.
\newblock
{\BBOQ}\APACrefatitle {Potential for added value to downscaled climate extremes
  over {K}orea by increased resolution of a regional climate model} {Potential
  for added value to downscaled climate extremes over {K}orea by increased
  resolution of a regional climate model}.{\BBCQ}
\newblock
\APACjournalVolNumPages{Theoretical and Applied
  Climatology}{117}{3-4}{667--677}.
\newblock
\begin{APACrefDOI} \doi{10.1007/s00704-013-1034-6} \end{APACrefDOI}
\PrintBackRefs{\CurrentBib}

\bibitem [\protect \citeauthoryear {%
Leung%
\ \BBA {} Qian%
}{%
Leung%
\ \BBA {} Qian%
}{%
{\protect \APACyear {2003}}%
}]{%
leung2003sensitivity}
\APACinsertmetastar {%
leung2003sensitivity}%
\begin{APACrefauthors}%
Leung, L\BPBI R.%
\BCBT {}\ \BBA {} Qian, Y.%
\end{APACrefauthors}%
\unskip\
\newblock
\APACrefYearMonthDay{2003}{}{}.
\newblock
{\BBOQ}\APACrefatitle {The sensitivity of precipitation and snowpack
  simulations to model resolution via nesting in regions of complex terrain}
  {The sensitivity of precipitation and snowpack simulations to model
  resolution via nesting in regions of complex terrain}.{\BBCQ}
\newblock
\APACjournalVolNumPages{Journal of Hydrometeorology}{4}{6}{1025--1043}.
\newblock
\begin{APACrefDOI} \doi{10.1175/1525-7541(2003)004<1025:TSOPAS>2.0.CO;2}
  \end{APACrefDOI}
\PrintBackRefs{\CurrentBib}

\bibitem [\protect \citeauthoryear {%
Maga{\~n}a%
, Amador%
\BCBL {}\ \BBA {} Medina%
}{%
Maga{\~n}a%
\ \protect \BOthers {.}}{%
{\protect \APACyear {1999}}%
}]{%
magana1999midsummer}
\APACinsertmetastar {%
magana1999midsummer}%
\begin{APACrefauthors}%
Maga{\~n}a, V.%
, Amador, J\BPBI A.%
\BCBL {}\ \BBA {} Medina, S.%
\end{APACrefauthors}%
\unskip\
\newblock
\APACrefYearMonthDay{1999}{}{}.
\newblock
{\BBOQ}\APACrefatitle {The midsummer drought over Mexico and Central America}
  {The midsummer drought over mexico and central america}.{\BBCQ}
\newblock
\APACjournalVolNumPages{Journal of Climate}{12}{6}{1577--1588}.
\newblock
\begin{APACrefDOI} \doi{10.1175/1520-0442(1999)012<1577:TMDOMA>2.0.CO;2}
  \end{APACrefDOI}
\PrintBackRefs{\CurrentBib}

\bibitem [\protect \citeauthoryear {%
Rew%
\ \BBA {} Davis%
}{%
Rew%
\ \BBA {} Davis%
}{%
{\protect \APACyear {1990}}%
}]{%
rew1990netcdf}
\APACinsertmetastar {%
rew1990netcdf}%
\begin{APACrefauthors}%
Rew, R.%
\BCBT {}\ \BBA {} Davis, G.%
\end{APACrefauthors}%
\unskip\
\newblock
\APACrefYearMonthDay{1990}{}{}.
\newblock
{\BBOQ}\APACrefatitle {{NetCDF}: {A}n interface for scientific data access}
  {{NetCDF}: {A}n interface for scientific data access}.{\BBCQ}
\newblock
\APACjournalVolNumPages{IEEE Computer Graphics and
  Applications}{10}{4}{76--82}.
\newblock
\begin{APACrefDOI} \doi{10.1109/38.56302} \end{APACrefDOI}
\PrintBackRefs{\CurrentBib}

\bibitem [\protect \citeauthoryear {%
Salath{\'e}%
, Leung%
, Qian%
\BCBL {}\ \BBA {} Zhang%
}{%
Salath{\'e}%
\ \protect \BOthers {.}}{%
{\protect \APACyear {2010}}%
}]{%
salathe2010regional}
\APACinsertmetastar {%
salathe2010regional}%
\begin{APACrefauthors}%
Salath{\'e}, E\BPBI P.%
, Leung, L\BPBI R.%
, Qian, Y.%
\BCBL {}\ \BBA {} Zhang, Y.%
\end{APACrefauthors}%
\unskip\
\newblock
\APACrefYearMonthDay{2010}{}{}.
\newblock
{\BBOQ}\APACrefatitle {Regional climate model projections for the State of
  Washington} {Regional climate model projections for the state of
  washington}.{\BBCQ}
\newblock
\APACjournalVolNumPages{Climatic Change}{102}{1}{51--75}.
\newblock
\begin{APACrefDOI} \doi{10.1007/s10584-010-9849-y} \end{APACrefDOI}
\PrintBackRefs{\CurrentBib}

\bibitem [\protect \citeauthoryear {%
Small%
, De~Szoeke%
\BCBL {}\ \BBA {} Xie%
}{%
Small%
\ \protect \BOthers {.}}{%
{\protect \APACyear {2007}}%
}]{%
small2007central}
\APACinsertmetastar {%
small2007central}%
\begin{APACrefauthors}%
Small, R\BPBI J\BPBI O.%
, De~Szoeke, S\BPBI P.%
\BCBL {}\ \BBA {} Xie, S\BHBI P.%
\end{APACrefauthors}%
\unskip\
\newblock
\APACrefYearMonthDay{2007}{}{}.
\newblock
{\BBOQ}\APACrefatitle {The Central American midsummer drought: regional aspects
  and large-scale forcing} {The central american midsummer drought: regional
  aspects and large-scale forcing}.{\BBCQ}
\newblock
\APACjournalVolNumPages{Journal of Climate}{20}{19}{4853--4873}.
\newblock
\begin{APACrefDOI} \doi{10.1175/JCLI4261.1} \end{APACrefDOI}
\PrintBackRefs{\CurrentBib}

\bibitem [\protect \citeauthoryear {%
Wallace%
\ \BBA {} Hobbs%
}{%
Wallace%
\ \BBA {} Hobbs%
}{%
{\protect \APACyear {2006}}%
}]{%
wallace2006atmospheric}
\APACinsertmetastar {%
wallace2006atmospheric}%
\begin{APACrefauthors}%
Wallace, J\BPBI M.%
\BCBT {}\ \BBA {} Hobbs, P\BPBI V.%
\end{APACrefauthors}%
\unskip\
\newblock
\APACrefYear{2006}.
\newblock
\APACrefbtitle {Atmospheric science: {A}n introductory survey} {Atmospheric
  science: {A}n introductory survey}\ (\PrintOrdinal{2nd. ed.}\ \BEd,
  \BVOL~92).
\newblock
\APACaddressPublisher{}{London: Academic press}.
\PrintBackRefs{\CurrentBib}

\bibitem [\protect \citeauthoryear {%
Wang%
\ \protect \BOthers {.}}{%
Wang%
\ \protect \BOthers {.}}{%
{\protect \APACyear {2004}}%
}]{%
wang2004regional}
\APACinsertmetastar {%
wang2004regional}%
\begin{APACrefauthors}%
Wang, Y.%
, Leung, L\BPBI R.%
, McGregor, J\BPBI L.%
, Lee, D\BHBI K.%
, Wang, W\BHBI C.%
, Ding, Y.%
\BCBL {}\ \BBA {} Kimura, F.%
\end{APACrefauthors}%
\unskip\
\newblock
\APACrefYearMonthDay{2004}{}{}.
\newblock
{\BBOQ}\APACrefatitle {Regional climate modeling: {P}rogress, challenges, and
  prospects} {Regional climate modeling: {P}rogress, challenges, and
  prospects}.{\BBCQ}
\newblock
\APACjournalVolNumPages{Journal of the Meteorological Society of
  Japan}{82}{6}{1599--1628}.
\newblock
\begin{APACrefDOI} \doi{10.2151/jmsj.82.1599} \end{APACrefDOI}
\PrintBackRefs{\CurrentBib}

\bibitem [\protect \citeauthoryear {%
Zender%
}{%
Zender%
}{%
{\protect \APACyear {2008}}%
}]{%
zender2008analysis}
\APACinsertmetastar {%
zender2008analysis}%
\begin{APACrefauthors}%
Zender, C\BPBI S.%
\end{APACrefauthors}%
\unskip\
\newblock
\APACrefYearMonthDay{2008}{}{}.
\newblock
{\BBOQ}\APACrefatitle {Analysis of self-describing gridded geoscience data with
  netCDF Operators (NCO)} {Analysis of self-describing gridded geoscience data
  with netcdf operators (nco)}.{\BBCQ}
\newblock
\APACjournalVolNumPages{Environmental Modelling \&
  Software}{23}{10}{1338--1342}.
\newblock
\begin{APACrefDOI} \doi{10.1016/j.envsoft.2008.03.004} \end{APACrefDOI}
\PrintBackRefs{\CurrentBib}

\end{thebibliography}


\end{document}